\documentclass[onecolumn, 12pt]{aastex631}

\usepackage{amsmath}
\usepackage{graphicx}
\usepackage{natbib}
\usepackage{hyperref}
\usepackage{cancel}
\usepackage{xcolor}
\usepackage{amssymb}

\shorttitle{The RAR from a 2VT-motivated framework}
\shortauthors{Dantas, Ribeiro \& Capelato}

\definecolor{r2}{rgb}{0.9,0.0,0.0}

\begin{document}

\title{The Radial Acceleration Relation in Galaxies and Clusters \\ from a Two-Component, Virial-Motivated Framework}

\correspondingauthor{Christine C. Dantas}
\email{christine.dantas@inpe.br}

\author[0000-0002-2833-2520]{Christine C. Dantas}
\affiliation{Divisão de Astrofísica, Instituto Nacional de Pesquisas Espaciais, São José dos Campos, 1
2227-010, SP, Brazil}

\author[0000-0002-3813-2565]{André L. B. Ribeiro}
\affiliation{Departamento de Ciências Exatas e Tecnológicas, Universidade Estadual de Santa Cruz, Ilhéus, 45650-000, BA, Brazil}
\email{albr@uesc.br}

\author[0000-0002-3664-599X]{Hugo V. Capelato}
\affiliation{Núcleo de Astrofísica Teórica (NAT), Universidade Cidade de São Paulo, São Paulo, 01506-000, SP, Brazil}
\email{hcapelato@gmail.com}

\begin{abstract}

We present a comparative analysis of acceleration data for gravitational systems drawn from multiple observational sources, including: (i) early-type galaxies (ETGs) \citep{Lel17}; (ii) Brightest Cluster Galaxies (BCGs) and galaxy clusters \citep{Tian20,Tian24}; and (iii) weak gravitational lensing of isolated galaxies \citep{Bro21,Mis24}. These data are interpreted within a framework motivated by the {\it Two-component Virial Theorem} (2VT), which defines a global baryon--dark matter (DM) coupling and sets a characteristic acceleration scale. This baseline is complemented by two models that account for the main empirical features of the {\it Radial Acceleration Relation} (RAR) over a broad range of masses and accelerations. The {\it Constant-Interaction Approximation Model} (CIA) reproduces the observed RAR trends for ETGs, BCGs, and galaxy clusters. It extends earlier results \citep{Dan00,Dan18}, and accounts for both the small intrinsic scatter and the emergence of a characteristic acceleration scale. At the very low accelerations probed by weak-lensing data (below accelerations of order $10^{-14}~\mathrm{m\,s^{-2}}$), however, this model breaks down. In this regime, the {\it Virial-Motivated Interaction Model} (VIM) incorporates the radial structure of the baryon--DM interaction through a local, radius-dependent contribution to the acceleration. Taken together, the 2VT (global scale), the CIA and the VIM provide a physically motivated framework that captures the main empirical features of the RAR.
\end{abstract}


\section{Introduction} \label{sec:int}

Dark matter (DM) is any form of matter inferred only through its gravitational effects on other bodies. In the standard $\Lambda$CDM cosmological paradigm (composed of dark energy, represented by the cosmological constant, $\Lambda$, cold dark matter, CDM, and baryonic matter and radiation), small primordial fluctuations grow by gravitational instability and collapse hierarchically into DM halos that host the baryonic matter (galaxies and galaxy clusters). Currently, the large-scale structure of the Universe can be explained by this framework with reasonable accuracy \citep{Planck2018,Abbott2022}.

At galaxy scales, however, there have been tensions between $\Lambda$CDM predictions and some observational inferences. N-body simulations originally produced centrally cuspy density profiles \citep[e.g.,][]{Nav97,Springel2008}, whereas rotation-curve analyses of dwarfs and low surface brightness galaxies favor shallower inner slopes or approximately constant-density cores \citep[e.g.,][]{deBlok2010,Oh2015,Read2017}. Additional small-scale discrepancies noted in the literature include the ``missing satellites'' and ``too-big-to-fail'' problems \citep[e.g.,][]{Klypin1999,Moore1999,BoylanKolchin2011}. Hydrodynamical simulations with stellar feedback have, however, addressed many of these issues with varying degrees of success \citep[e.g.,][]{Pontzen2012,BenitezLlambay2019}. 

An important empirical evidence is the \textit{Radial Acceleration Relation} (RAR). Using the SPARC database of high-quality rotation curves and resolved baryonic mass models, \citet{McC16} and \citet{Lel17} demonstrated that the total observed acceleration $g_{\rm obs}$ correlates tightly with the acceleration predicted by the observed baryons $g_{\rm B}$, with small intrinsic scatter and a characteristic acceleration scale $a_M \sim 10^{-10}\,[{\rm m\,s^{-2}}]$ (hereon referred as the {\it Milgrom’s acceleration scale}). This relation implies that, for the population of disk galaxies, the total radial acceleration is almost uniquely specified by the baryonic contribution.

Several lines of work indicate that emergent RAR-like behavior can arise in CDM due to the combined influence of baryonic physics and the structural response of halos to baryon condensation and feedback. However, high-resolution hydrodynamic cosmological simulations have provided different results: some reproduce RAR-like trends with small scatter, while others deviate depending on feedback prescriptions and stellar-to-halo mass relations \citep[e.g.,][]{Ludlow2017,Navarro2017}. Analytic arguments have also connected the RAR to halo concentration, assembly history, and baryon-driven relaxation processes \citep[e.g.,][]{Keller2017,Desmond2017}, implying that RAR need not require new gravity if galaxy formation produces the right baryon–DM couplings.

The RAR has been probed with complementary methods and galaxy types, beyond rotation curves. Weak-lensing measurements extend tests to larger radii and galaxy samples \citep{Bro21}, while satellite galaxies and low-mass systems provide tests of RAR universality \citep{Paranjape2021}. At cluster scales, the situation is more complex, as brightest cluster galaxies (BCGs) and cluster mass distributions probe accelerations well below and above $a_M$, and some recent cluster analyses suggest departures from the galaxy RAR \citep{Tian20,Tian24}.

The RAR has three crucial empirical properties that require explanation: (i) its small scatter, (ii) the presence of the characteristic acceleration scale, $\sim a_M$, below which apparent mass discrepancies grow, and (iii) the fact that the baryonic distribution predicts the dynamical acceleration even in regions where baryons are not as dominant. These properties were anticipated qualitatively by Milgrom’s {\it Modified Newtonian Dynamics} (MOND) \citep{Mil83,Milgrom2014} and subsequent relativistic extensions (e.g., \citealt{Bekenstein2004}), in which $a_M$ is a fundamental constant and the RAR is a manifestation of modified dynamics rather than an unseen (DM) gravitational potential.

Evidences for the universality of the RAR across galaxy types and environments provide an important empirical signature, in order to address whether it favors modified-gravity frameworks or tightly coupled baryon–halo interactions. Such a universality could also be the result of an emergent relation, with measurable dependence on formation history or environment, which nevertheless would require a detailed development of baryon--DM coupled physics along the cosmological evolution, for its explanation. 

Our line of research proposes the application of the {\it Two-component Virial Theorem} (2VT) as a viable model for explaining the scaling relations of all virialized stellar systems \citep{Dan00}. Subsequently from our first investigations, we have also studied the predictions of the global scaling of the 2VT for the RAR, which reproduced reasonable fits for galaxies, and specially predicting the low acceleration horizontal tail observed in dSphs \citep{Dan18}. The 2VT expresses a global correction to the usual virial theorem for the steady state equilibrium of a baryonic system that is embedded in a larger DM halo. This alters the dynamics of the combined system, so that in addition to the baryonic self-potential, the baryonic component is also subjected to a baryon-weighted potential including the DM contribution.

In the present study, we examine the RAR across a range of gravitational systems, including early-type galaxies (ETGs) \citep{Lel17}, Brightest Cluster Galaxies (BCGs), and galaxy clusters \citep{Tian20,Tian24}, as well as weak-lensing measurements of isolated galaxies \citep{Bro21,Mis24}. We interpret these data within a two-component, virial-motivated framework that connects the dynamics of baryons and DM across different mass and acceleration regimes. Within this framework, the 2VT provides a global constraint on the baryon--DM coupling, defined at the baryonic gravitational radius, and sets a characteristic acceleration scale for each system. In the radial range probed by the observed RAR, where the DM contribution varies smoothly within the baryonic region, this global scale can be approximated as an effective, nearly constant interaction term. This leads to a {\it Constant-Interaction Approximation Model} (CIA), which captures the main features of the RAR for ETGs, BCGs, and galaxy clusters, extending earlier applications of the 2VT \citep{Dan00,Dan18}.
At the very low accelerations probed by weak-lensing data (below $\sim 10^{-14}~\mathrm{m\,s^{-2}}$), however, the assumption of a radially constant interaction term is no longer valid, as the radial structure of the DM halo becomes observationally relevant. This motivates a more general formulation, in which the interaction between baryons and DM is treated explicitly as a radius-dependent quantity. We therefore introduce the {\it Virial-Motivated Interaction Model} (VIM), which defines a local interaction acceleration that extends the baryon–DM coupling underlying the 2VT to a radially resolved description.  These two models are therefore complementary descriptions of the baryon--DM coupling at different levels.

This paper is structured as follows. In Sec. \ref{sec:2VT}, we provide an overview of the 2VT, the CIA, and the VIM. In Sec. \ref{sec:Met} we present the methodology (datasets and best-fit procedures) and in Sec. \ref{sec:results} we analyze the CIA and the VIM predictions against the data. In Sec. \ref{sec:disc}, we discuss our results and provide a detailed interpretation of the mechanisms behind the RAR in our  framework, and conclude in Sec. \ref{sec:conc}. In App. \ref{APP}, we provide a derivation of the original 2VT, the CIA, and the VIM. The notation used throughout this work is summarized in Table~\ref{tab:notation} in that appendix.

\section{An overview of the CIA and VIM models} \label{sec:2VT}

The virial theorem provides a global constraint on the dynamical equilibrium of self-gravitating systems. However in the case of a baryonic component embedded in a DM halo, the relevant formulation should be a two-component virial theorem which was first proposed by Limber in order to model a slightly different situation (\citealt{Lim59}; see also  \citealt{Dan00,Dan18}). The two-component virial theorem (2VT: cf App. \ref{ssec:2VT_3D}) extends the standard scalar virial theorem (1VT: cf App. \ref{ssec:1VT}) by including the gravitational interaction between baryons and DM.

The 2VT is inherently a global relation, defined at the baryonic gravitational radius, $r_B$, and does not by itself prescribe the radial dependence of the gravitational acceleration. Instead, it defines a characteristic acceleration scale associated with the baryon--DM coupling, set by the global structural properties of the system. This distinction is essential when comparing with the RAR, which is constructed from measurements at multiple radii within galaxies and clusters. The 2VT is the basis for the introduction of the CIA and VIM models. In this section we provide a qualitative exposition of the CIA and the VIM models, with details provided in the Apps. \ref{ssec:CIA} and \ref{APP_VIM}, respectively.

\subsection{The Constant-Interaction Approximation Model (CIA)} \label{ssec:2VT_CIA}

This model is motivated by the global scale set by the 2VT. Over the radial range probed by the RAR, where the DM contribution varies smoothly within the baryonic region, the 2VT scale can be approximated as an effective constant interaction term ($g_{\rm int}$, cf. App. \ref{ssec:CIA}), leading to this model. The term $g_{\rm int}$ is a system-dependent constant set by the global baryon--DM coupling.  This approximation is intended to apply, within individual systems, over the radial range sampled by the data where the DM contribution varies smoothly, while variations in $g_{\rm int}$ across different systems reflect differences in their global structure, such as the distribution of baryons and the central density of the DM halo. It reproduces the observed RAR trends across ETGs, BCGs, and galaxy clusters, and naturally accounts for both its small intrinsic scatter and the emergence of a characteristic acceleration scale. It is important to note that CIA {\it predicts} a horizontal tail at low baryonic accelerations, accommodating particularly well the dSph data \citep{Lel17}. The logarithmic representation of the CIA and its limiting behavior is summarized in the App. \ref{ssec:log_CIA}.

We emphasize that the CIA model is not intended to predict the detailed radial dependence of the acceleration within individual systems. Rather, it provides an effective description of the global baryon–DM coupling that constrains the overall normalization and asymptotic behavior of the acceleration. In this sense, the CIA captures system-level properties that manifest in the RAR when considering ensembles of measurements across different radii and across different systems. 

In the radial range typically probed by the RAR, the DM contribution within the baryonic region varies smoothly and can be approximated, to first order, as an effective constant. Under this condition, the global interaction scale defined by the 2VT can be used as an effective approximation for radially resolved measurements through the CIA model (cf. App. \ref{ssec:CIA}), in which the observed acceleration is written as
\begin{equation}
g_{\rm CIA}(r_{\mathrm{obs}}) \simeq g_B(r_{\mathrm{obs}}) + g_{\rm int}, \label{eq:CIA}
\end{equation}
where $g_{\rm int}$ is a system-dependent constant set by the global baryon--DM coupling  and $g_B$ is the radius-dependent baryonic contribution. This approximation applies within individual systems, in radial regimes where the DM contribution varies smoothly,  over the radial range sampled by the data, while variations in $g_{\rm int}$ across different systems reflect differences in their global structure, such as the distribution of baryons and the central density of the DM halo. In the intermediate-to-low acceleration regime, the global baryon–DM interaction term dominates the dynamics. This naturally leads to the emergence of the horizontal tail of the RAR (e.g., as populated by dSphs). Radius-dependent baryonic contributions enter as subdominant corrections, while at high accelerations the relation approaches the Newtonian relation, $g_B= G M_B(<r_{obs})/r^2_{obs}$ (cf. App. \ref{ssec:CIA}).

Within this approximation, the effective interaction term can be expressed in terms of the quantities entering the 2VT. In particular, one may write
\begin{equation}
g_{\rm int} \equiv \mathcal{R}\, g_D(r_B),\label{eq:CIA_int}
\end{equation}
where $g_D(r_B)$ is the characteristic DM acceleration evaluated at the baryonic gravitational radius, and $\mathcal{R}$ (Eq.~\ref{eq:R_param}) is a dimensionless parameter that encodes the geometrically baryon-weighted contribution of the DM halo.

\subsection{The Virial-Motivated Interaction Model (VIM)} \label{ssec:VIM}

This model  can be viewed as a radial generalization of the CIA. The CIA provides a leading-order, scale-independent approximation to the baryon--DM coupling within the baryonic region, whereas the VIM generalizes this description by explicitly incorporating the radial structure of the interaction (c.f. App. \ref{APP_VIM}). The correction term in the VIM is therefore radius-dependent, in contrast with the global interaction scale encoded in $g_{\rm int}$, and becomes essential in regimes where the constant-interaction approximation breaks down.

At the very low accelerations probed by weak-lensing data, the constant-interaction approximation breaks down. For this regime, the VIM model incorporates the radial structure of the baryon--DM interaction through a local, radius-dependent contribution to the acceleration.  It provides a good fit to weak-lensing data in the case of a NFW DM profile. Although the VIM is not itself a virial theorem and does not encode equilibrium conditions, it preserves the physical scale set by the 2VT, while providing a force-based, radially resolved description suitable for comparison with local acceleration data.

Indeed, the RAR for the new data based on gravitational lensing of isolated galaxies does not show such a horizontal trend at low accelerations \citep{Bro21,Mis24}, contrary to the observed horizontal tail in the dSph data \citep{Lel17}. These systems are therefore not well described by a constant interaction term, indicating that the CIA is no longer an adequate approximation in this regime and motivating the introduction of the VIM, given by:
\begin{equation}
g_{\rm VIM} (r_{\mathrm{obs}}) = g_B (r_{\mathrm{obs}}) + g_{BD}(r_{\mathrm{obs}}),\label{eq:g_ext_COPY}
\end{equation}
\noindent where $g_{BD}$ refers to the numerically computed correction to the acceleration. In this sense, the VIM generalizes the CIA, in which the interaction term is no longer approximated as constant but retains its explicit dependence on radius.

The free parameters of the VIM are the usual DM mass density profile parameters, namely, the density normalization, $\rho_0$, and scale radius, $r_s$ (c.f. App. \ref{APP_profiles} for the profiles used in this work). A further free parameter, $\alpha$, relating the baryonic and DM density profiles, is also considered (Eq. \ref{eq:alpha}). The procedure for obtaining the VIM best-fit to the dataset is presented in Sec. \ref{ssec:2VT_ext_best_fit}.

\section{Methodology} \label{sec:Met}

\subsection{Summary of the dataset} \label{ssec:dataset}

In this section we briefly present the new radial acceleration dataset used for obtaining and fitting the RAR, which we add to a previous dataset reported in \citet{Dan18}. The dataset includes a variety of observational sources, now incorporating new measurements for early-type galaxies (ETGs), brightest cluster galaxies (BCGs), intracluster radial accelerations in clusters of galaxies and accelerations inferred from weak lensing of isolated galaxies. The dataset used in \citet{Dan18} is also used for the present 2VT-RAR fits, which includes late-type (LTGs) and dwarf spheroidals (dSphs), from \citet{McC16, Lel16, Lel17}. We list the new dataset as follows:

\begin{itemize} 
\item{Rotational ETGs at two different locations, labeled as {\it inner}: measurements at the effective radius; and {\it outer}: measurements via HI \citep{Lel17};}
\item{X-ray ETGs at various radii \citep{Lel17};}
\item{Central BCGs of 20 clusters from CLASH \citep{Tian20};}
\item{50 BCGs from the MaNGA survey \citep{Tian24};}
\item{Intra-cluster regions from CLASH, i.e., accelerations measured at 4 distances (namely, at $r = \{ 100, 200, 400, 600 \}$ kpc from the center of each cluster  \citep{Tian20,Tian24};}
\item{Isolated galaxies: accelerations measured via weak gravitational lensing against a background of galaxies.  This sample is based on the Kilo-Degree Survey (KiDS-1000; \citealt{Jon13,Kui19}) combined with the Galaxy and Mass Assembly (GAMA) survey \citep{Dri11}, originally analyzed by \citet{Bro17,Bro21}. 
More recently, \citet{Mis24} reanalyzed the same lensing dataset, applying a deprojection procedure to infer three-dimensional accelerations and adopting an alternative reconstruction of the observed radial acceleration, $g_{\mathrm{obs}}$.  While the underlying lensing measurements are identical to those of \citet{Bro21}, the methodology of \citet{Mis24} yields a deprojected RAR that is directly comparable to three-dimensional dynamical models.}
\end{itemize} 

\subsection{Empirical models for comparison with the CIA} \label{ssec:empirical}

We briefly describe some empirical models in the literature that are relevant to be compared to the 2VT (accelerations are all in $[\mathrm{m\,s^{-2}}]$):

The initially proposed RAR model given in \citet{McC16},
\begin{equation}
\mathcal{F}_a(g_B) \;\equiv\;  g_{\rm obs}(g_B) \;=\; 
\frac{g_B}{1 - \exp\!\left(-\sqrt{\dfrac{g_B}{a_{M}}}\,\right)} , \label{eq:McG16}
\end{equation}
\noindent with $a_M = 1.20 \times 10^{-10} ~[\mathrm{m\,s^{-2}}]
$ (Milgrom's acceleration scale).

A modification of the above equation was suggested in order to fit the horizontal trend of low acceleration data presented by dSphs \citep{Lel17},
\begin{equation}
\mathcal{F}_b(g_B) \;\equiv\;  g_{\rm obs}(g_B) \;=\; 
\frac{g_B}{1 - \exp\!\left(-\sqrt{\dfrac{g_B}{g_{\dagger}}}\,\right)} 
+ \hat{g} \, \exp\!\left(-\sqrt{\dfrac{g_B\,g_{\dagger}}{\hat{g}^2}}\,\right) , \label{eq:Lel17}
\end{equation}
\noindent with the best fit parameters: $g_{\dagger} = (1.1 \pm 0.1) \times 10 ^{-10} ~[\mathrm{m\,s^{-2}}]
$  and $\hat{g} = (9.2 \pm 0.2) \times 10 ^{-12} ~[\mathrm{m\,s^{-2}}]
$.

Another empirical RAR relation, deviating from the above RAR models at galaxy scales, was obtained for $20$ massive galaxy clusters in the Cluster Lensing And Supernova survey with Hubble (CLASH), as described in \citet{Tian20}:
\begin{equation}
g_{\rm obs}(g_B)  \simeq \sqrt{g_B \; g_{\ddagger}}
\; \Longrightarrow \; 
\log [g_{\rm obs}(g_B) ] \simeq  0.5 \left [ \log  (g_B) + \log (g_{\ddagger}) \right ],
\label{eq:Tian20}
\end{equation}
\noindent which corresponds to a power-law relation of slope 1/2 in logarithmic space, with $g_{\ddagger} = (2.0\pm 0.1) \times 10^{-9}  ~[\mathrm{m\,s^{-2}}]
$, that is, a characteristic acceleration scale that is larger than that found in the RAR for galaxies. 
The combined $50$ MaNGA BCGs, $20$ CLASH BCGs, and $64$ clusters data, confirmed the previously found {\it distinct RAR in cluster scales} \citep{Tian24}:
\begin{equation}
\mathcal{F}_c (g_B) \;\equiv\;  \log [g_{\rm obs}(g_B) ] \;=\; 
0.52^{+0.01}_{-0.01} \log (g_B) - 4.19 ^{+0.15}_{-0.15}. \label{eq:Tian24}
\end{equation}

\subsection{Mathematical comparison of the empirical models with the CIA } \label{ssec:Emp_CIA}

A useful way to contrast the CIA with empirical RAR formulations is to examine its behavior in the transition regime between baryon-dominated and interaction-dominated dynamics. In the CIA, this regime is defined by $g_B(r_{\mathrm{obs}}) \sim g_{\rm int}$. In this limit, the CIA relation reduces to
$g_{\rm CIA} \simeq 2 g_B$, which in logarithmic form becomes $\log g_{\rm CIA} \simeq \log g_B + 0.3$.
This corresponds to a linear relation of unit slope, offset by $\sim 0.3$ dex above the Newtonian expectation. This straight line provides a convenient local approximation to the CIA curve in the transition region. 

However, that linear behavior does not reflect the intrinsic scaling of the model. The true logarithmic slope of the CIA relation at $g_B \sim g_{\rm int}$ is
\begin{equation}
\frac{d \log g_{\rm CIA}}{d \log g_B} \simeq \frac{1}{2},
\end{equation}
indicating that the CIA retains a curved relation in this regime. This curvature arises from the additive form $g_{\rm obs} = g_B + g_{\rm int}$, and reflects the gradual transition between the two dynamical contributions.

By contrast, the empirical RAR proposed by \citet{McC16} (Eq.~\ref{eq:McG16}) does not admit such a decomposition. Its transition around $g_B \sim a_M$ is intrinsically smooth and governed by a nonlinear rescaling of $g_B$, leading asymptotically to $g_{\rm obs} \propto \sqrt{a_M g_B}$ at low accelerations. Although both frameworks exhibit a local logarithmic slope of order $1/2$ in the transition region, their origins are different: in the CIA, this behavior emerges from the balance of two additive components, whereas in the McGaugh relation it reflects an intrinsic scaling property of the functional form.
A similar distinction applies to the modified relation proposed by \citet{Lel17} (Eq.~\ref{eq:Lel17}), which reproduces the low-acceleration flattening through an additional exponential term.

\subsection{Procedure for obtaining the best-fit to the CIA} \label{ssec:2VT_best_fit}

We briefly describe here the numerical procedure adopted to obtain the best-fit parameters of the CIA. We start by highlighting a slight change of notation and procedure relatively to our previous work \citep{Dan18}, where we fixed  $g_D \equiv a_M = 1.20 \times 10^{-10} ~[m/s^2]$ (Milgrom's acceleration scale). In this case, the best-fit parameter was quoted as the value of the dimensionless parameter $\mathcal{R}$ (c.f. Eqs. \ref{eq:R_param}). 

In the present paper, we adopt a revised parameterization consistent with the CIA, so that the best-fit parameter is quoted in terms of a single dimensionless quantity, $\mathcal{G}$ (see below). We do not assume any fixed value for $g_D$, as done previously. Instead, we express the effective interaction term in units of $a_M$, writing
\begin{equation}
g_{\rm int} = \mathcal{G}\, a_M. \label{eq:mathcalG}
\end{equation}
For the numerical implementation, one may equivalently write $g_D = f \cdot a_M$, with $f$ dimensionless, so that: $\mathcal{R} g_D = \mathcal{R} f\, a_M = \mathcal{G} a_M \Longrightarrow  \mathcal{G} \equiv \mathcal{R} f$. In order to connect with our previous work, we recall that, by fixing there $g_D = a_M$, the best-fit was quoted as $\mathcal{R}  g_D = 1.82 \times 10^{-11} ~[m/s^2]$, leading to $\mathcal{R}=0.152$ (c.f. Fig. 2 in \citealt{Dan18}). In the present work, this translates to $\mathcal{G} = 0.152 = \mathcal{R} f$.

Therefore, different CIA curves are determined by different values of $\mathcal{G}$, which directly sets the amplitude of the interaction term and controls the transition from the baryon-dominated regime to the regime in which the interaction term becomes significant ($g_B \lesssim g_{\rm int}$). In logarithmic space, this produces the characteristic curvature and eventual flattening toward the low-acceleration limit.
Hence, for any given system or dataset, the importance of the interaction term is determined relative to the range of $g_B$ probed by the observations. There remains a degeneracy in the decomposition of $\mathcal{G}$ into $\mathcal{R}$ and $f$, as these quantities are not independently constrained. However, the CIA depends only on the combined parameter $\mathcal{G}$, which provides a compact and directly testable description of the baryon--DM coupling.

The CIA best-fit parameters for the various cases reported here, were obtained through non-linear least-squares minimization using the \texttt{curve\_fit} routine from the \texttt{scipy.optimize} package \citep{SciPy20,Python09}. This method adjusts the model function to the data by minimizing the weighted sum of squared residuals, where the weights are given by the observational uncertainties. The fitting accounts for the reported errors via the \texttt{absolute\_sigma=True} option. 

\subsection{Procedure for obtaining the best-fit to the VIM} \label{ssec:2VT_ext_best_fit}

For a given DM density profile (Navarro--Frenk--White, Burkert, cored isothermal, and Moore; see App.~\ref{APP_profiles}), we computed the local baryon–DM interaction acceleration $g_{BD}(r)$ directly from its 3D definition (c.f. Eq. \ref{eq:gBD_local}). The DM density $\rho_D(r)$ was evaluated at the
radius of interest, while the enclosed DM mass was computed numerically by direct quadrature.
The VIM model predictions were evaluated at the effective radii reported by \citet{Bro21}, and compared to the deprojected (3D) radial acceleration relation measured by
\cite{Mis24}. 

The parameters of the model (namely, the central DM density $\rho_0$, the scale radius $r_s$, and the dimensionless interaction parameter $\alpha$; c.f. Eq. \ref{eq:alpha}) were determined by a direct $\chi^2$ minimization against the RAR data of \citet{Mis24}. No parameters are fixed \emph{a priori}; in particular, $\alpha$ is allowed to vary freely and is constrained by the data. Given the limited number of data points, we restrict the analysis to the best-fit solution and do not attempt to derive formal confidence bands, which would not be statistically meaningful in this regime.

In the numerical implementation, the baryonic mass scale entering the interaction term is fixed to
\( M_B = 10^{10.96}\,M_\odot \), corresponding to the highest-mass bin of the sample in \cite{Bro21}.  In the local formulation adopted here, the specific value of $M_B$ only sets an overall normalization of the interaction term. Since $\alpha$ is treated as a free parameter and determined by the fit to the RAR data, any change in the reference baryonic mass scale can be absorbed by a corresponding rescaling of $\alpha$, leaving the shape of the predicted relation of different halo profiles unchanged. The choice of the highest-mass bin therefore serves as a convenient normalization reference and does not introduce an additional physical scale into the model.

\section{Results} \label{sec:results}

\subsection{The CIA and the RAR} \label{ssec:2VT_results}

\begin{figure}[b]
\centering
\includegraphics[width=0.9\linewidth]{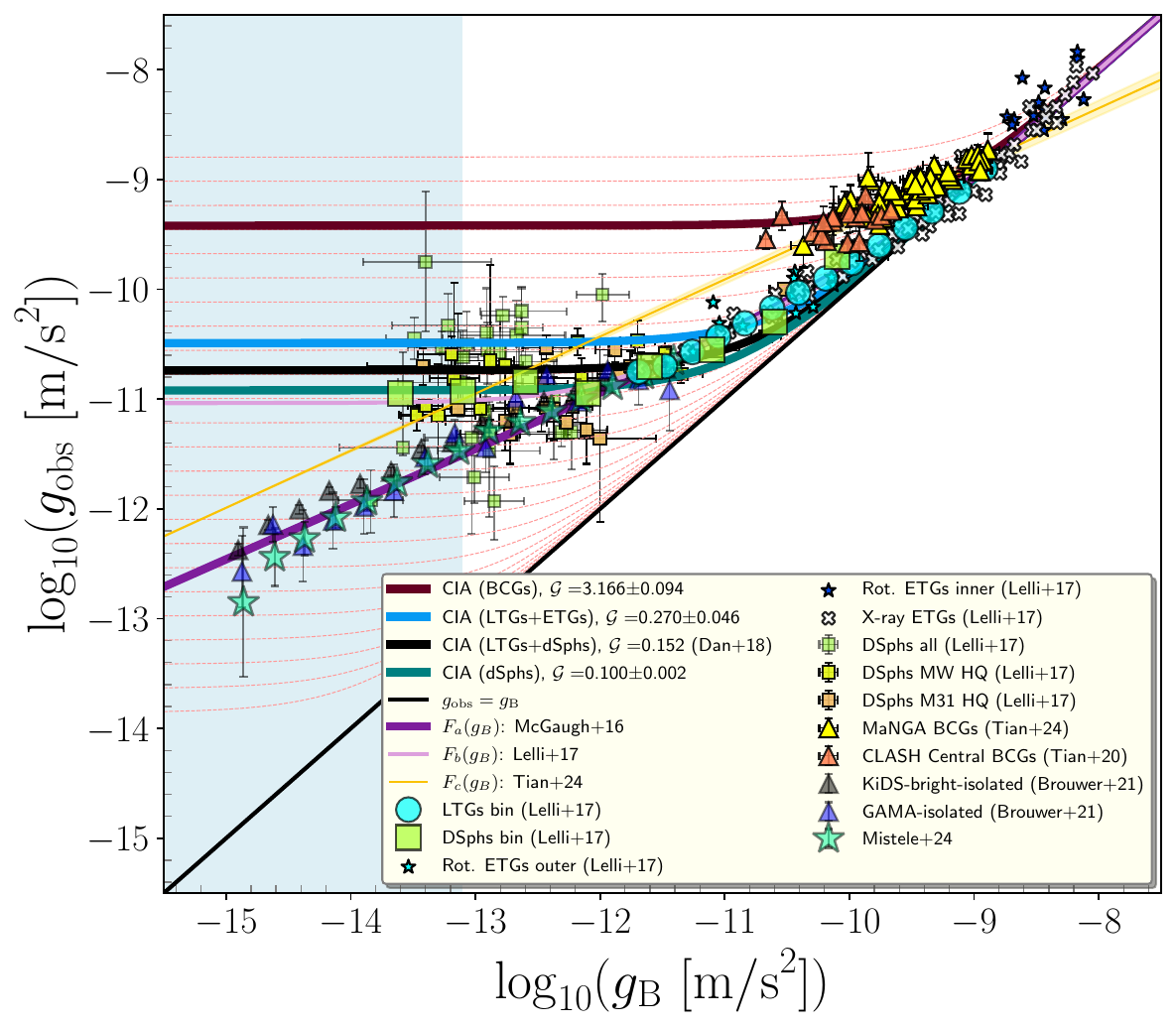}
\caption{The CIA and the RAR for galaxies (ETGs, LTGs, dSphs, and BCGs), including the empirical model curves for comparison. The light blue region in the very low baryon acceleration range signals an uncertainty in the photometric KiDS redshifts as described by \citet{Bro21}. The VIM curves for this lower acceleration regime are treated separately in Sec. \ref{ssec:2VT_ext}.}
\label{fig:New_2VT}
\end{figure}

\begin{figure}[htbp]
\centering
\includegraphics[width=0.495\linewidth]{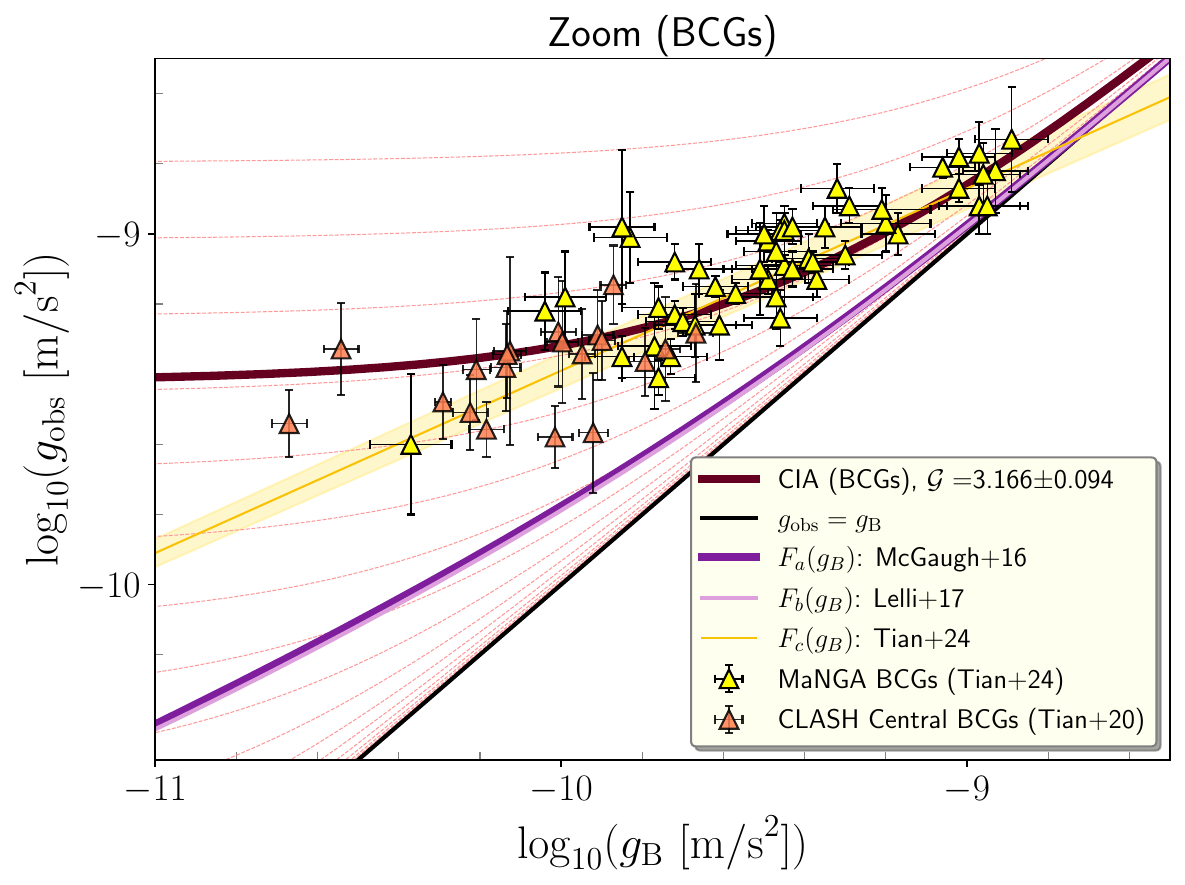}
\includegraphics[width=0.495\linewidth]{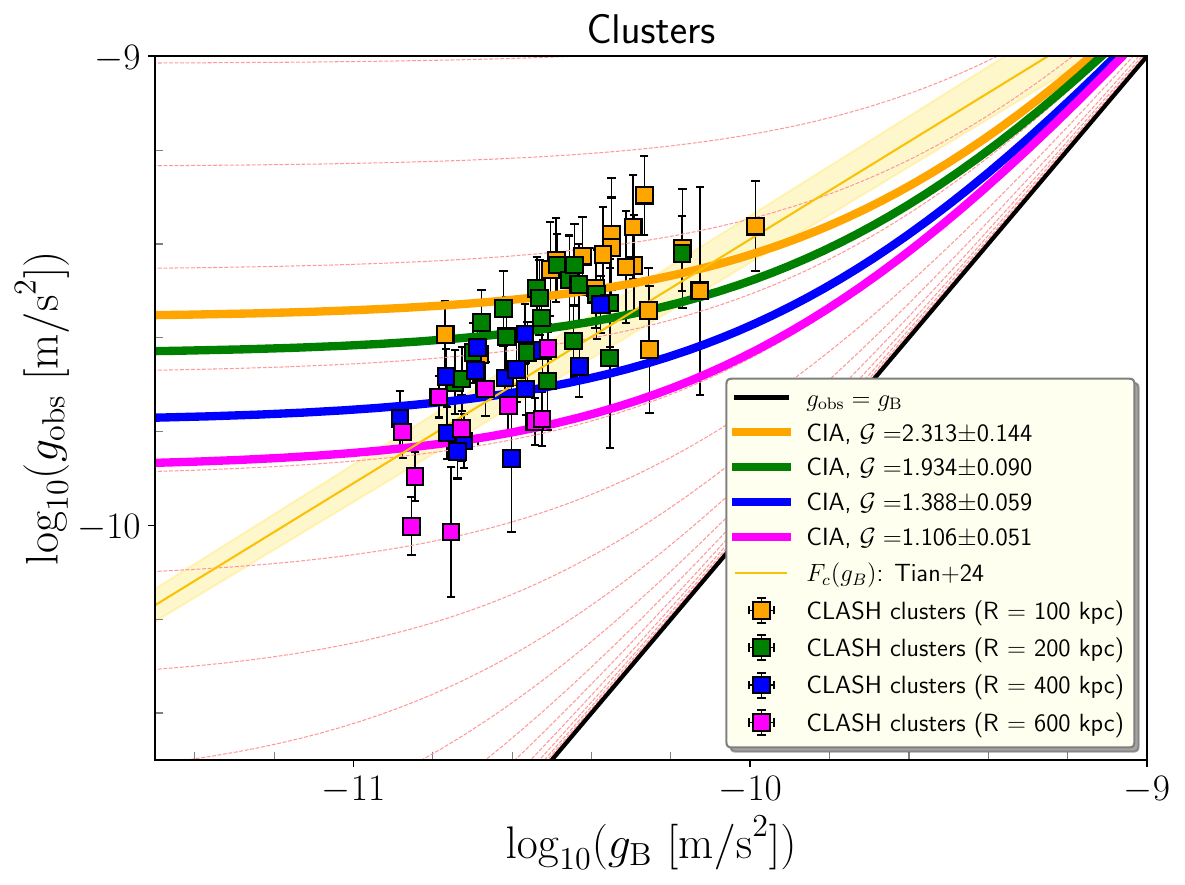}
\caption{RAR curves and data symbols are the same as in Fig. \ref{fig:New_2VT}. {\it Left panel:} Zoom-in for the RAR of BCGs only. {\it Right panel:} The RAR for the 20 CLASH clusters, considering $4$ distinct cluster-centric radii $R$, given in the legend. Distinct CIA curves for each $R$ are shown, with their best-fit $\mathcal{G}$ values. The best-fit curve given by \cite{Tian24} is also indicated.}
\label{fig:New_2VT_BCGs_Clusters}
\end{figure}

In Fig. \ref{fig:New_2VT}, we plot the RAR for the complete dataset (c.f. Sec. \ref{ssec:dataset}), except for the intracluster data, which are presented separately in the right panel of Fig. \ref{fig:New_2VT_BCGs_Clusters} (to be discussed below). All empirical curves listed in Sec. \ref{ssec:empirical} are shown for comparison with the CIA best-fits in terms of the parameter $\mathcal{G}$. The very low acceleration data for isolated galaxies \citep{Bro21,Mis24} are shown for comparison; however, the VIM curves for this data are treated separately in Sec. \ref{ssec:2VT_ext}, for a detailed discussion and comparison with the model by \cite{McC16}.

The CIA clearly accommodates most of the dataset, with different (yet, somewhat close) values of $\mathcal{G}$ for the case of galaxies (ETGs, LTGs, dSphs), but with a significantly higher $\mathcal{G}$ value for BCGs. The CIA fails to reproduce the isolated galaxy lenses dataset of \cite{Bro21} in the very low baryon acceleration range ($\log(g_B) \lesssim -13$). As already mentioned, for this dataset, we analyze the performance of the VIM, with results discussed in Sec. \ref{ssec:2VT_ext}. 

The left panel of Fig. \ref{fig:New_2VT_BCGs_Clusters} shows a zoom-in of Fig. \ref{fig:New_2VT} for the BCGs only, in order to clearly display the CIA best-fit curve, which accounts for the functional behavior of the RAR for this dataset, including the observed horizontal tendency of $\log(g_{\rm obs})$ for lower values of $\log(g_B)$. In the right panel of Fig. \ref{fig:New_2VT_BCGs_Clusters}, we plot the RAR for the CLASH intracluster accelerations evaluated at four distinct distances from the respective cluster centers. In this case, the CIA fits were performed separately for each of the four subsets. These fits also accommodate the data, with a noticeable systematic decrease of $\mathcal{G}$ for larger cluster-centric distances. As progressively larger radii are probed, the baryonic component becomes less extended relative to the aperture, so that the effective interaction term becomes less significant (the departure from the baryon-dominated regime then occurs at progressively smaller $\log(g_B)$ values).

\begin{figure}[htbp]
\centering
\includegraphics[width=0.495\linewidth]{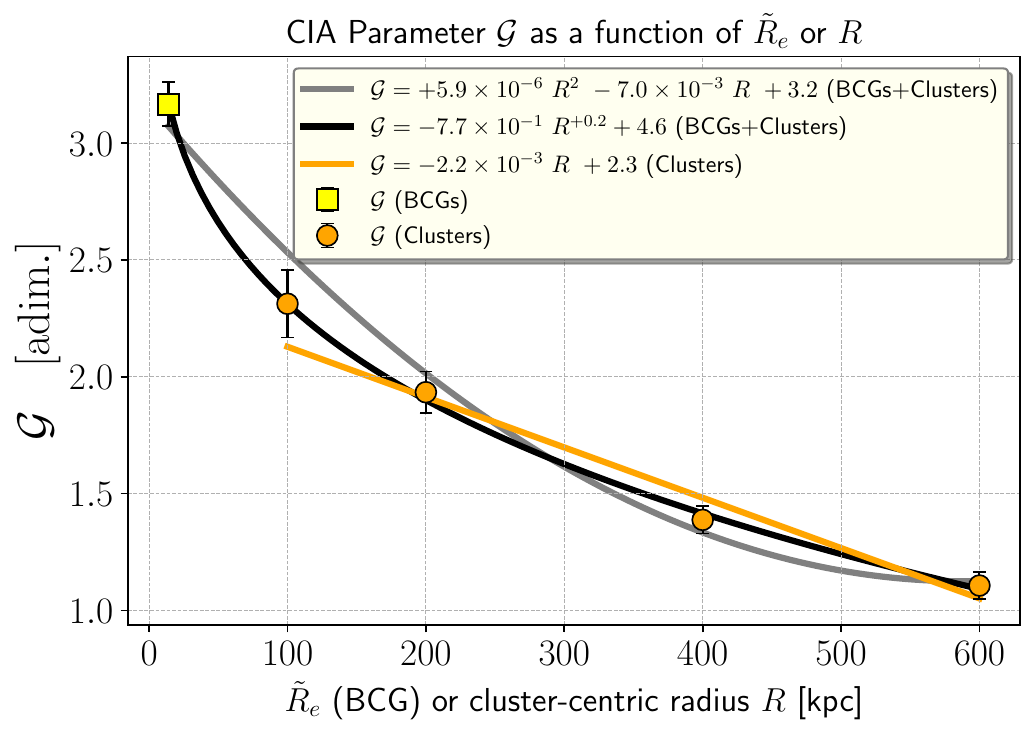}
\includegraphics[width=0.495\linewidth]{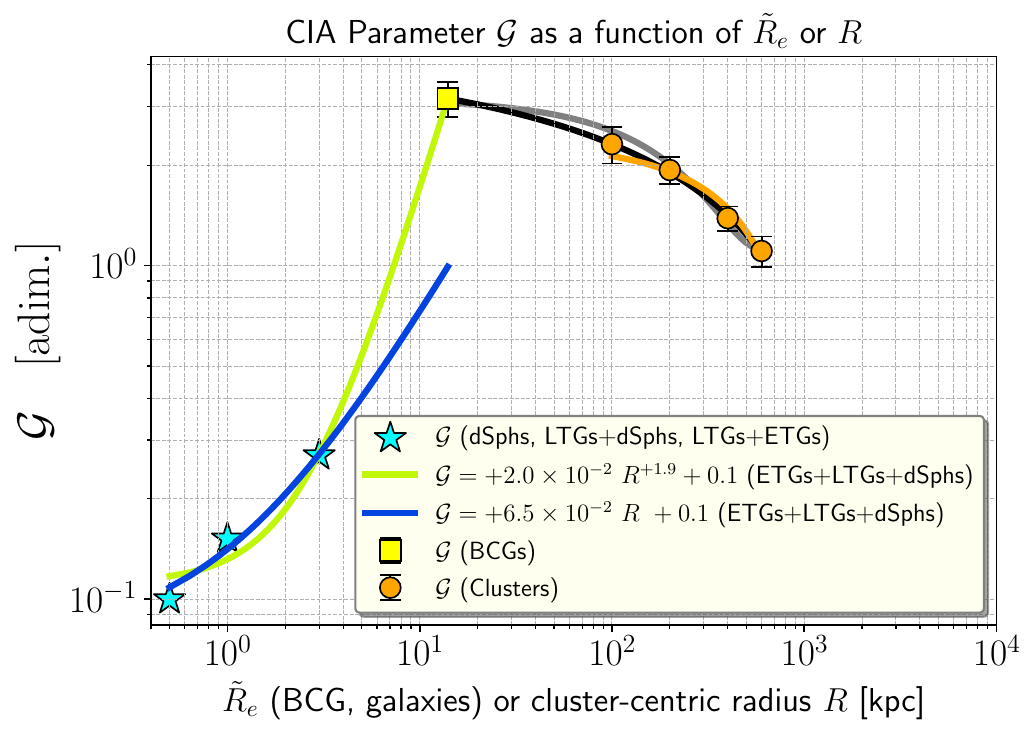}
\caption{{\it Left panel:} Dependence of the CIA best-fit parameter $\mathcal{G}$ on the reference effective radius, $\tilde{R}_e$ (BCGs), or on the cluster-centric distance, $R$ (clusters). Two quasi-linear models are shown in black and grey lines (fit considering BCGs and intracluster data). A linear model was fit for the intracluster data only (orange line). {\it Right panel:} Same as the left panel, now including the CIA best-fit parameters, $\mathcal{G}$, for galaxies (dSphs, LTGs+dSphs, LTGs+ETGs; blue line). A distinct behavior between both sets (BCGs and clusters vs. galaxies) is clearly seen. The green line shows the functional requirement for connecting these two sets.}
\label{fig:Merge_Figs_4_5}
\end{figure}

In the left panel of Fig. \ref{fig:Merge_Figs_4_5} we show $\mathcal{G}(R)$ for the dataset of \citet{Tian20,Tian24}, where for BCGs we use the reference effective radius, $\tilde{R}_e = 14$ kpc, which is the approximate average for those systems. Two sets of fits are shown, one using all data (BCGs and intracluster regions), with two possible quasi-linear fits, and the other for intracluster data only, with a linear fit. In the right panel of Fig. \ref{fig:Merge_Figs_4_5}, we consider the entire dataset. We use typical average values of $\tilde{R}_e = \{0.5, 1.0, 3.0\}$ kpc for dSphs, LTGs+dSphs, and LTGs+ETGs, respectively. Galaxies follow a different regime in terms of $\mathcal{G}(R)$, with an inverse behavior in $R$ compared to BCGs and intracluster datasets. As larger values of $\tilde{R}_e$ are probed in galaxies, the baryonic distribution appears more extended within the effective radius relative to smaller systems such as dSphs. However, most LTGs and ETGs are located at higher values of $g_B$ than dSphs. Therefore, the interaction term plays a more prominent role in the case of dSphs. We also show in the right panel of Fig. \ref{fig:Merge_Figs_4_5} a ``virtual connection'' between both regimes, which would require $\mathcal{G}(R) \sim R^2$ for galaxies to reach the $\mathcal{G}$ values of BCGs. This is purely illustrative, as a detailed physical modeling would be required to assess such a transition, which lies beyond the scope of the present work.

The results presented in this section indicate that the CIA parameter, $\mathcal{G}$, is not universal, but varies systematically with the physical scale of the system. While galaxies (from dSphs to ETGs) and BCGs are each well described by approximately constant values of $\mathcal{G}$ within their respective baryonic extents, galaxy clusters exhibit a clear radial dependence, such that $\mathcal{G} = \mathcal{G}(R)$, where $R$ denotes the cluster-centric radius. This behavior is illustrated by the fact that cluster data are not fitted by a single CIA relation, but instead by a family of CIA curves, each corresponding to a different radial aperture. The inferred values of $\mathcal{G}$ decrease systematically with increasing radius, reflecting the changing structural and dynamical conditions across the cluster environment. In contrast, the approximately constant $\mathcal{G}$ values found for galaxies and BCGs indicate that, within their baryonic regions, the DM contribution can be effectively represented by a single interaction scale.

\subsection{The VIM and the RAR} \label{ssec:2VT_ext}

In this section we analyze the RAR under the VIM. The data by \citet{Mis24} were derived from the same galaxy--galaxy lensing observations originally analyzed by \citet{Bro21}, but employed a different methodology to reconstruct the observed acceleration, based on a 3D (deprojected) treatment of the lensing signal. The VIM best-fits were obtained exclusively by fitting the dataset by \citet{Mis24}. The data measurements from \citet{Bro21}, based on the KiDS and GAMA surveys, are shown in the figures solely for reference and comparison with earlier projected RAR determinations. They are not included in the fitting procedure and do not enter the determination of the VIM parameters.

\begin{figure}[t]
\centering
\includegraphics[width=0.495\linewidth]{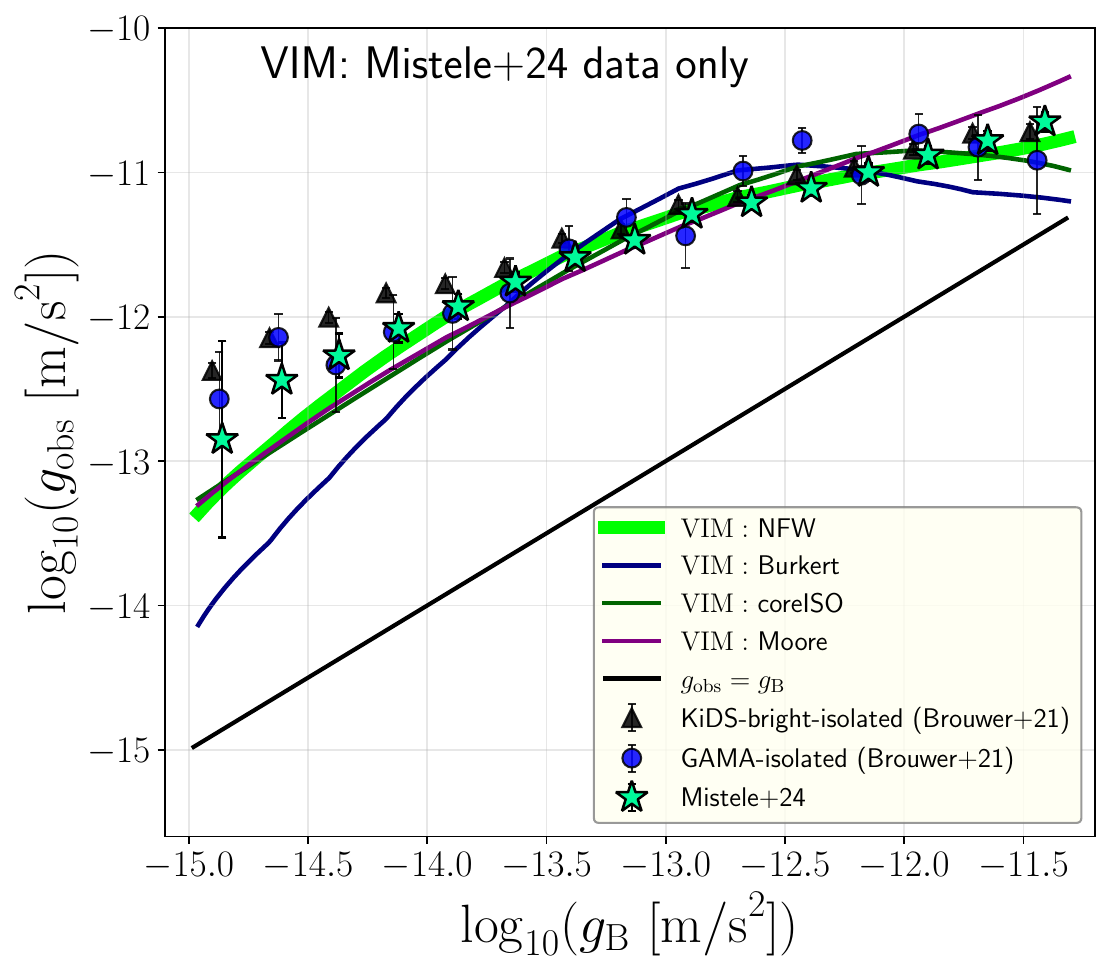}
\includegraphics[width=0.495\linewidth]{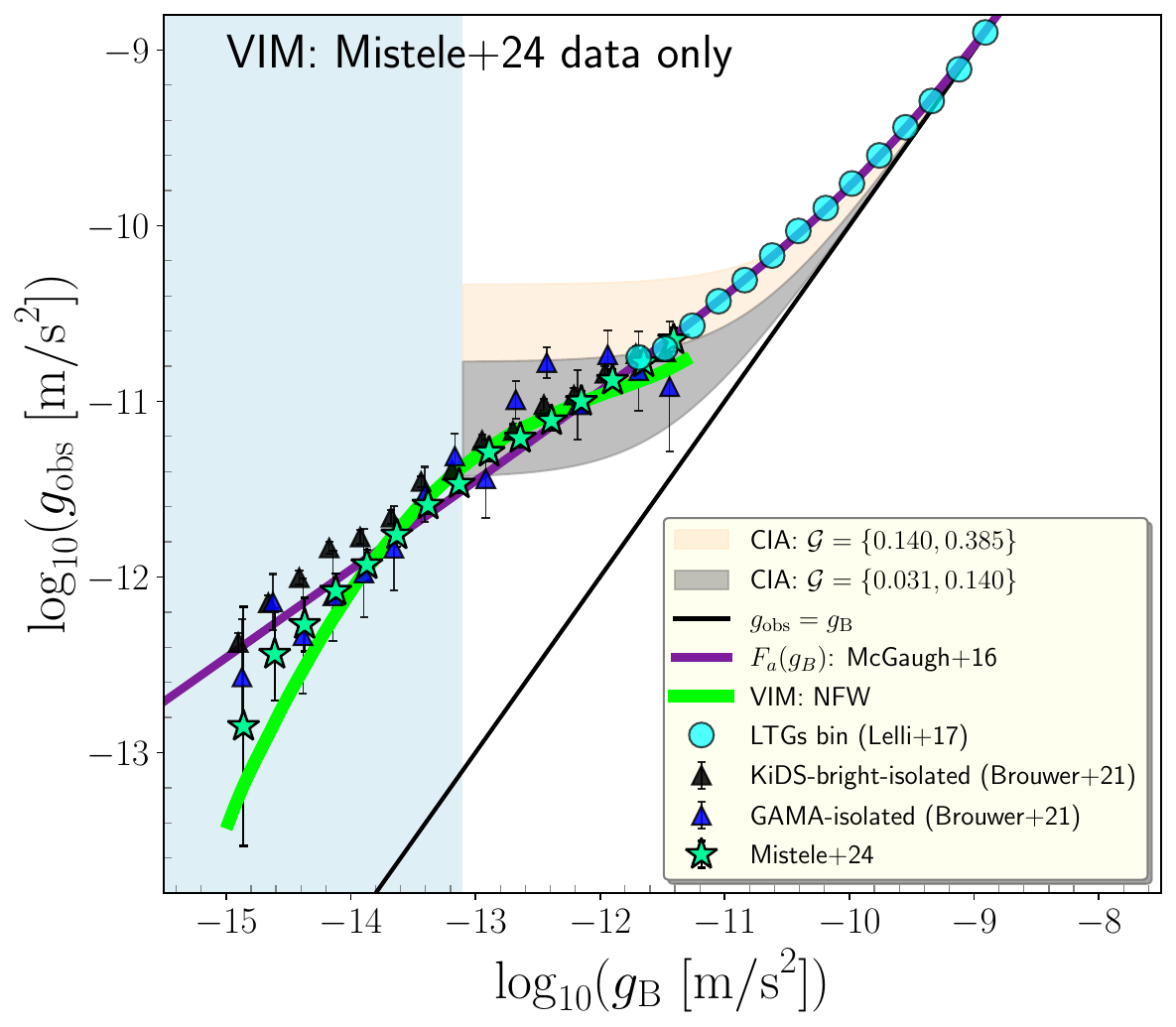}
\caption{
{\it Left panel:} Best-fit of the VIM model for isolated galaxy lenses, obtained by fitting the deprojected RAR data of \citet{Mis24} using four different DM density profiles (indicated in the legend). The isolated-galaxy lensing measurements of \citet{Bro21}, based on the KiDS and GAMA surveys, are shown for reference only and are not included in the fitting procedure. 
{\it Right panel:} Zoom-out view of the same VIM best-fit relations over a wider range of baryonic accelerations. Binned LTG data are included as a reference for the high-acceleration regime.
}
\label{fig:2VT_ext}
\end{figure}

Figure~\ref{fig:2VT_ext} (left panel) shows the resulting best-fit VIM curves for the four DM profiles (Navarro–Frenk–White, Burkert, cored isothermal, and Moore; c.f. App. \ref{APP_profiles}), compared with the isolated-galaxy lensing data.  The best-fit parameters are shown in Tab. \ref{tab:2VText_fits}. Note that the fitted density normalizations ($\rho_0$)   should be interpreted as effective parameters constrained by the limited acceleration range probed by the RAR, rather than as global halo densities. Their intermediate values are larger than those inferred in the original 2VT analysis at higher accelerations (c.f. \citealt{Dan00}), yet smaller than canonical cosmological halo normalizations.

The parameter $\alpha$ introduced in the VIM (c.f. Eqs. \ref{eq:alphaR} and \ref{eq:alpha}) represents a local proportionality between baryonic and DM densities. However, its best-fit value varies considerably among the halo profiles considered (Tab. \ref{tab:2VText_fits}), ranging from $\alpha \sim 10^{-4}$ for the Moore profile to $\alpha \sim 1$ for the cored models. This variation reflects a degeneracy in fitting DM halo profiles to limited radial data: cuspier profiles (NFW, Moore) require a smaller baryon--DM coupling normalization to produce a comparable interaction acceleration at the probed radii, whereas cored profiles, with their shallower inner density, require a larger $\alpha$ to match the same observed radial acceleration, $g_{\mathrm{obs}}$.

The VIM treats $\alpha$ as an effective coupling parameter constrained by the RAR data, not a scale-free proportionality between the intrinsic density profiles of baryons and DM. In a more complete physical model, $\alpha$ would be expected to vary with radius and to correlate with the formation history of the galaxy, reflecting processes such as adiabatic contraction, feedback-driven core formation, or the hierarchical assembly of the halo. From this perspective, the best-fit $\alpha$ values derived here should be interpreted as weighted averages over the radial range effectively probed by the weak-lensing RAR. Cosmological simulations that simultaneously track the RAR and the underlying halo properties \citep[e.g.,][]{Ludlow2017, Navarro2017} could provide valuable priors on the expected range and radial dependence of such an effective coupling parameter within the $\Lambda$CDM framework.

Among the profiles considered, the NFW halo yields
the best acceptable fit, while simultaneously reproducing the mild downturn observed at the lowest baryonic accelerations.  This behavior indicates that the observed RAR favors a halo structure of intermediate cuspiness, for which both the local DM density and the baryon–DM interaction contribute comparably to the total acceleration. In contrast, the Burkert, cored isothermal, and Moore profiles yield significantly poorer fits, requiring extreme values of the $\alpha$ parameter to compensate for their inner density structure.

In the right panel of Fig.~\ref{fig:2VT_ext}, we present a zoom-out view of the same best-fit relations, extending to higher accelerations for an overall visualization of the RAR space. Binned LTG data are included as a reference for the high-acceleration regime, where the relation gradually approaches the one-to-one expectation $g_{\mathrm{obs}} = g_{\mathrm{B}}$. While the original 2VT predicts a constant asymptotic acceleration at low $g_{\mathrm{B}}$, it is unable to account for the observed curvature of the RAR below $g_{\mathrm{B}} \lesssim 10^{-13}\,\mathrm{m\,s^{-2}}$. The VIM, through its local interaction term, naturally reproduces the observed downward bending of the relation in this regime and remains compatible with the empirical RAR parametrization of \citet{McC16} down to $g_{\mathrm{B}} \lesssim 10^{-14}\,\mathrm{m\,s^{-2}}$.

\begin{table}[t]
\centering
\caption{Best-fit parameters of the VIM to the deprojected RAR data of Mistele et al.~(2024),
for different dark–matter density profiles.}
\label{tab:2VText_fits}
\begin{tabular}{lcccccc}
\hline\hline
Profile
& $\rho_0$ 
& $r_s$
& $\alpha$
& $\chi^2$
& dof
& $\chi^2/{\rm dof}$ \\
& [$\mathrm{M_\odot\,kpc^{-3}}$]
& [kpc]
& 
& 
& 
&  \\
\hline
NFW        & $5.00\times10^{2}$ & $0.656$ & $0.0347$ & $20.12$ & $12$ & $1.68$ \\
Burkert    & $1.27\times10^{3}$ & $0.164$ & $1.0000$ & $245.92$ & $12$ & $20.49$ \\
coreISO    & $2.06\times10^{3}$ & $0.062$ & $1.0000$ & $74.67$ & $12$ & $6.22$ \\
Moore      & $5.30\times10^{2}$ & $2.047$ & $0.0001$ & $65.30$ & $12$ & $5.44$ \\
\hline
\end{tabular}
\end{table}

\section{Discussion}  \label{sec:disc}

We discuss our results relatively to the main empirical features of the RAR.

\begin{itemize}

\item{{\it A small intrinsic scatter.} In the case of an approximately constant parameter $\mathcal{G}$ (Eq.\ref{eq:mathcalG}), within a given population (e.g., LTGs, ETGs, dSphs), the CIA predicts a one-to-one relation between $g_{\rm obs}$ and $g_B$, with an additive constant interaction term ($g_{\rm int}$), which could explain the observed tight scatter of the RAR. In the very low-acceleration regime, the VIM preserves this behavior provided the additional interaction term remains subdominant or varies smoothly with radius, such that it does not introduce significant additional scatter. Residual scatter may then be attributed to secondary variations in baryonic structure, halo properties, or environment, which translate into small population-dependent deviations in $\mathcal{G}$ or in the normalization parameter $\alpha$.}

\item{{\it A characteristic acceleration scale $a_M$.} The CIA interaction term, $g_{\rm int} = \mathcal{G} a_M$ (Eq.\ref{eq:mathcalG}), defines a transition below which the interaction contribution becomes significant relative to $g_B$. For $g_B \gg g_{\rm int}$, baryons dominate and $g_{\rm obs} \approx g_B$ (the Newtonian or 1VT regime), while for $g_B \lesssim g_{\rm int}$ the interaction term produces the curvature and eventual flattening of the relation. In this sense, the acceleration scale $a_M$ emerges naturally through the normalization of the interaction term, reflecting the underlying baryon--DM coupling rather than representing a fundamental constant. At lower accelerations  ($\log g_B \lesssim -14$), beyond the scale $a_M$, the VIM is able to produce a downturn and provides a detailed shape of the RAR in the DM-dominated regime in the case of a NFW profile. The other DM profiles investigated (Burkert, cored isothermal, and Moore profiles) provide poorer fits. However, all the VIM best-fits present a curvature relative to the empirical RAR model of \cite{McC16} (c.f. Eq. \ref{eq:McG16}).}

\item{{\it The ability of the baryonic distribution to predict the total dynamical acceleration even in DM-dominated regions.} In the case of the CIA,  even when the baryonic mass is small compared to the DM mass, the effective interaction term remains tied to the global baryonic structure (e.g., through its scale or concentration), ensuring that the total acceleration remains correlated with the baryonic distribution. The VIM reinforces this coupling in the low-acceleration limit, where the additional interaction term depends on the local baryonic mass scale and halo structure. This seems to provide a relative baryon-driven imprint on the observed acceleration, even deep in the DM-dominated regime.}

\item{{\it The RAR for clusters}. Within our framework, the apparent radial acceleration relation observed for clusters \citep{Tian20,Tian24}, characterized by a power-law behavior $g_{\rm obs} \propto g_B^{1/2}$ (cf. Eq. \ref{eq:Tian20}), can be interpreted as an emergent relation. Rather than reflecting a fundamental modification of the dynamical law, this behavior could arise from the continuous variation of the effective interaction term, $g_{\rm int}(R) \propto \mathcal{G}(R)$, combined with the radial dependence of the baryonic acceleration. In other words, the cluster RAR corresponds to a projection of a family of CIA relations with scale-dependent interaction strength. }

\end{itemize}

\clearpage

Some additional comments on our results are now briefly considered.

\begin{itemize}

\item {{\it The distinct behavior of the CIA parameter $\mathcal{G}(R)$ in galaxies and clusters.} 
For galaxies, $\mathcal{G}$ is observed to decrease with increasing effective radius (from dSphs to ETGs), whereas for BCGs and intracluster regions, $\mathcal{G}$ increases with scale.
This inversion suggests a change in the dominant structural factor controlling the baryon--DM coupling.
In the relatively isolated, low-mass systems (dSphs), the baryonic component is deeply embedded within a DM halo whose central density is high relative to the baryonic scale radius, leading to a proportionally larger interaction term (higher $\mathcal{G}$). As one moves to more massive, extended galaxies, the baryonic distribution becomes less deeply embedded, and $\mathcal{G}$ declines. In contrast, within the dense environment of a galaxy cluster, the baryonic component of a BCG or the intracluster light is embedded in a DM halo whose overall mass and concentration grow substantially with radius over the scales probed, causing $\mathcal{G}$ to rise. A detailed modeling of this transition would require accounting for the environmental dependence of halo concentration, the effects of tidal stripping and mergers, and the possible contribution of a diffuse intra-cluster baryonic component not associated with a single galaxy. While a full treatment is beyond the scope of this work, the empirical trends identified here offer a clear target for future theoretical studies aiming to connect the RAR to the cosmic evolution of DM halos across mass and environmental density.
}
\item {{\it The downward deviation from the asymptotic flattening.} At the lowest baryonic accelerations ($\log g_B \lesssim -14$), the best-fit VIM prediction for all profiles exhibits a mild but
systematic downward deviation from the asymptotic flattening expected
from purely scale-free modifications of gravity. This behavior appears to
originate from the explicitly local nature of the baryon--DM
interaction term, $g_{BD}(r)$, which depends on both the enclosed
DM mass and the local DM density. As the baryonic
component probes increasingly large radii, the product
$\rho_D(r)\,M_D(<r)/r$ decreases,
leading to a reduction of the interaction contribution to the total
acceleration. The resulting downturn is therefore not an artifact of
the fitting procedure, but a structural feature of the
VIM formulation when applied locally. At this point, it remains unclear whether this downturn in the RAR reflects an intrinsic property of the lens systems or instead signals a limitation of the isolation criterion, an issue that warrants further investigation (c.f. discussion in \citealt{Mis24}). This behavior thus provides a nontrivial test of the predictions of the VIM.
}
\end{itemize}

\section{Conclusions} \label{sec:conc}

The dynamics of galaxies and clusters of galaxies provide a critical testing ground for alternative gravity models and extensions of the standard $\Lambda$CDM cosmology. A useful framework for studying these effects is the 2VT, which can accommodate baryonic and dark components simultaneously. We revisited the predictions of the 2VT considering new acceleration datasets available in the literature, and their radially resolved implementation through the CIA, as well as its extension to a fully local description via the VIM. Given new datasets, the present work provided an independent test of this framework relatively to our previous investigations \citep{Dan00,Dan18}, and a source of comparison with empirical models in the literature \citep{McC16,Lel17,Tian20}.

The present study confirmed the viability of the CIA and the VIM, in accounting for observed galactic and intracluster acceleration relations across different systems.  A key finding of our analysis is a variation of the CIA best-fit parameter $\mathcal{G}$ with the astrophysical environment. We demonstrated that $\mathcal{G}$ is considerably higher for BCGs and within cluster environments compared to isolated galaxies, and exhibits a distinct dependency on radial distance in intracluster regions. This environmental sensitivity of $\mathcal{G}$ may provide insights into the interplay between baryonic distribution and effective DM density. These results therefore support a unified picture in which the CIA provides a minimal, physically motivated description of the baryon--DM coupling across a wide range of systems, while the observed diversity of RAR behaviors reflects the scale dependence of the effective interaction parameter $\mathcal{G}$. In this sense, galaxies, BCGs, and clusters probe different regimes of the same underlying framework.

The CIA and the VIM, expressed in the acceleration space, closely resemble the observed RAR and, as such, are an outcome within $\Lambda$CDM, without the need for a specific/exotic baryon--DM microphysics, or modifications of gravity. The CIA provides a physically motivated account of the three key empirical properties of the RAR: the small scatter, the characteristic acceleration scale, and the predictive role of baryons in DM-dominated regions. The environmentally dependent nature of $\mathcal{G}$ may offer an indirect probe the fundamental nature of DM. By distinguishing how the parameters governing DM halo profiles ($ \rho_0, r_s$) are constrained across diverse environments and comparing these constraints with predictions from various DM candidates (e.g., Cold, Warm, Self-Interacting, or Fuzzy Dark Matter), future studies utilizing this framework can help to refine our understanding of the microscopic properties of the DM particle. The downturn in the RAR shown by the VIM is compatible with recent findings \citep{Mis24}, which nevertheless are under investigation. This behavior thus provides a further nontrivial test of the predictions of the VIM. Indeed, future larger and more precise datasets in the very low acceleration range will be instrumental not only for a more quantitative basis for differentiating the predictions of the CIA (and its VIM extension) against other models but also for breaking degeneracies and informing DM particle physics.

The predictive power of the framework presented here can be tested with forthcoming observational campaigns.  Larger samples of isolated galaxies with deep weak-lensing observations, expected from the Euclid Wide Survey, the Vera C. Rubin Observatory Legacy Survey of Space and Time (LSST), and the Nancy Grace Roman Space Telescope, will be important to confirm or refute the mild downturn in the RAR at the very low accelerations. Equally important, spatially resolved kinematics of low surface brightness galaxies and ultra-diffuse galaxies in group and cluster environments will allow us to map the transition in the $\mathcal{G}(R)$ relation across the full range of environments. This may provide tests to whether the CIA parameter is indeed a continuous function of halo mass, concentration, and environment. Such measurements will not only refine our understanding of the baryon–DM coupling but may also provide indirect constraints on the microphysical properties of DM, as different DM candidates (e.g., self-interacting, fuzzy, or warm DM) predict distinct inner halo structures and, consequently, distinct signatures in the low-acceleration RAR.

$~$ \\
$~$ \\
$~$ \\

\begin{acknowledgments}
We thank the referee for the constructive comments and suggestions, which have significantly improved the clarity and robustness of this work. We thank Yong Tian, Federico Lelli, and Margot Brouwer for making their data available. CCD thanks Brazilian Ministry of Science, Technology and Innovation, and the Brazilian Space Agency (AEB), which supported the present work under PO 20VB.0009.
ALBR thanks the support of CNPq, under grants 316317/2021-7 and 404160/2025-5.

\end{acknowledgments}

\software{Python (http://www.python.org),
          Matplotlib (http://dx.doi.org/10.1109/MCSE.2007.55),
	      Numpy (http://dx.doi.org/10.1038/s41586-020-2649-2),
          Scipy (http://dx.doi.org/10.1038/s41592-019-0686-2),
          Jupyter Notebook (http://dx.doi.org/10.3233/978-1-61499-649-1-87). }

\clearpage

\appendix

\section{The Two-Component, Virial-Motivated Frameworks} \label{APP}

In this Appendix, we provide a self-contained exposition of the Two-Component Virial Theorem (2VT) and the related, 2VT-motivated frameworks, namely, the Constant-Interaction Approximation (CIA) and the Virial-Motivated Interaction Model (VIM), clarifying several points and refining aspects of our previous formulations \citep{Dan00,Dan18}.  Throughout this work, we distinguish between global quantities defined at the baryonic gravitational radius (e.g., in the 2VT) and local quantities defined at arbitrary radii (e.g., in the CIA and VIM). The notation used throughout this work is summarized in Table~\ref{tab:notation}.

We also note that, in principle, the quantities defined within the virial framework could be related to observable accelerations if the gravitational radius and the full three-dimensional kinematics of the system were known. In practice, however, these quantities are not directly accessible, and are therefore treated here as formal acceleration scales characterizing the global dynamical state of the system.

\begin{table}[htbp]
\centering
\caption{Main symbols and notation used in this work}
\label{tab:notation}
\begin{tabular}{lll}
\hline\hline
Symbol & Units & Meaning \\
\hline

$r$ & kpc & Radius \\

$r_{\mathrm{obs}}$ & kpc & Observational radius \\

$r_B$ & kpc & Baryonic gravitational radius \\

$r_s$ & kpc & DM scale radius \\

$M_B(<r)$ & $M_\odot$ & Enclosed baryonic mass \\

$M_D(<r)$ & $M_\odot$ & Enclosed DM mass \\

$\rho_B(r)$ & $M_\odot$ kpc$^{-3}$ & Baryonic density \\

$\rho_D(r)$ & $M_\odot$ kpc$^{-3}$ & Dark matter density \\

$W_{BD}$ & -- & Interaction potential energy \\

$g_{\rm 1VT}(r_B)$ & m s$^{-2}$ & Global cceleration scale defined by 1VT\\

$g_{\rm 2VT}(r_B)$ & m s$^{-2}$ & Global acceleration scale defined by the 2VT \\

$g_{\rm vir,B}(r_B)$ & m s$^{-2}$ & Virial baryonic acceleration\\

$g_B(r)$ & m s$^{-2}$ & Newtonian baryonic acceleration ($GM_B(<r)/r^2$) \\

$g_D(r_B)$ & m s$^{-2}$ & Effective DM acceleration at $r_B$  \\ 

$g_{\rm int}$ & m s$^{-2}$ & Constant interaction term (CIA) \\

$g_{BD}(r)$ & m s$^{-2}$ & Local interaction term (VIM) \\

$g_{\rm obs}(r)$ & m s$^{-2}$ & Observed total acceleration 

(${\rm obs} \in \{ {\rm CIA}, {\rm VIM}, \mathcal{F}_a, \mathcal{F}_b, \mathcal{F}_c \} $) \\ 

$a_M$ & m s$^{-2}$ & Characteristic acceleration scale \\

$\mathcal{G}$ & -- & CIA parameter ($g_{\rm int}=\mathcal{G}a_M$) \\

$\mathcal{R}$ & -- & 2VT geometrical parameter \\

$\alpha$ & -- & Baryon--DM interaction parameter \\

\hline
\end{tabular}
\end{table}

\subsection{The One-Component Virial Theorem, 1VT} \label{ssec:1VT}

We begin with the (scalar) virial theorem (or here denoted {\it The One-Component Virial Theorem}, 1VT). Applied to a stationary self-gravitating system, it states that:
\begin{equation}
2K + W = 0, \label{eq:1VT}
\end{equation}
\noindent  where $K$ is the kinetic energy and $W$ is the potential energy of the system. This may be rewritten as:
\begin{equation}
\langle v^2 \rangle = {GM \over r_G}, \label{eq:1VT_v2}
\end{equation}
\noindent  where $\langle v^2 \rangle$ is the mean square velocity of the particles,  $G$ is the gravitational constant, $M$ is the total mass of the system, and $r_G$ is the gravitational radius, defined by:
\begin{equation}
r_G \equiv {GM^2 \over \mid W \mid}.  \label{eq:r_G}
\end{equation}
Note that Eq. \ref{eq:1VT_v2} does not require spherical symmetry, but only that the system is stationary (virialized) and that the potential energy is well-defined. Then, given the above expressions, the 1VT implies:
\begin{equation}
\langle v^2 \rangle = {\mid W \mid \over M}. \label{eq:1VT_v2_WM}
\end{equation}

On the other hand, for {\it spherically symmetric systems} (using Newton's shell theorem), the absolute value of the gravitational acceleration experienced by a test mass at a given radius, $r$, is given by :
\begin{equation}
g(r) = {GM(<r) \over r^2},    \label{eq:gr}
\end{equation}
\noindent where $M(<r)$ is the mass enclosed by a spherical volume of radius $r$. By comparing this expression (Eq. \ref{eq:gr}) with the 1VT (in the form of Eq. \ref{eq:1VT_v2}), the acceleration of a test mass placed at the gravitational radius of a spherically symmetric system under stationary equilibrium is then given by:

\begin{equation}
g_{\rm vir}(r_G) ={\langle v^2 \rangle \over r_G}.  \label{eq:grG}
\end{equation}
\noindent Therefore, if there was no DM, the 1VT in terms of (isotropic) radial accelerations would be the identity relation, with $r_B \equiv r_G$:
\begin{equation}
g_{\rm 1VT} (r_B) = g_{\rm vir, B}(r_B) = {\langle v^2 \rangle \over r_B}.  \label{eq:1VT_g}
\end{equation}

Note that the 1VT relation (Eq. \ref{eq:1VT_g}) in terms of accelerations is as a global identity evaluated at the gravitational radius of the system, which in the absence of DM is composed of baryons only.

A particular expression for the radial acceleration in a spherically symmetric system is obtained by assuming a constant density $\rho$ inside an arbitrary radius $r$ in Eq. \ref{eq:gr}, so that $M(<r) = {4 \pi \over 3} r^3 \rho $, which gives:
\begin{equation}
g(r) = {4 \pi G  \over 3} r \rho, ~~~~  \rho = {\rm const}. \label{eq:gr_rhoconst}
\end{equation}
\noindent Note that, contrary to the 1VT in terms of accelerations (Eq. \ref{eq:1VT_g}), the acceleration given by the previous expression is not restricted to the gravitational radius.

\subsection{The Two-component Virial Theorem, 2VT} \label{ssec:2VT_3D}

 Let $B$ refer to the baryonic component, and $D$ to the DM component. Then, the 2VT expression that is a counterpart to Eq. \ref{eq:1VT_v2_WM} is given by:	

\begin{equation}
\langle v^2 \rangle_{\rm 2VT}  \equiv {\mid W_{\rm 2VT}  \mid \over M_B(< r_G)} = {\mid W_B + W_{BD} \mid \over M_B(< r_G)}, \label{eq:2VT_v2_WM}
\end{equation}	
\noindent In what follows, $M_B$  will denote the total baryonic mass enclosed within the baryonic gravitational radius, $r_B \equiv r_G$; the (self-) gravitational potential energy of the baryonic component is
\begin{equation}
W_{B} = - G \int_0^\infty {\rho_B(r) M_B(<r) \over r} dV, \label{eq:WB}
\end{equation}
\noindent and the cross (interaction) gravitational potential energy is
\begin{equation}
W_{BD} = - G \int_0^\infty {\rho_B(r) M_D(<r) \over r} dV, \label{eq:WBD}
\end{equation}
\noindent with $dV = 4 \pi r^2 dr$ in the integrals above.
Considering the derivation of the 2VT  (c.f. \citealt{Dan00,Dan18}), we assumed that the DM halo is more spatially extended than the baryonic component, and having a not too steep density profile within the
region containing the baryonic component. Then we approximate: $M_D(<r) \approx {4 \pi \over 3} r^3 \rho_{0,D}$, where $\rho_{0,D}$ is the the mean density of the DM halo within the region containing the baryonic component, a region here considered to be a spherical volume within the gravitational radius of the baryonic component, $r_B$. Then we approximate $W_{BD}$ to:

\begin{equation} \label{eq:WBDappr}
\begin{split}
W_{BD}  & \approx - {4 \pi G\over 3} \rho_{0,D} \int_0^{r_B}  4\pi r^4 \rho_B(r) dr  \\
        & = - {4 \pi G\over 3} \rho_{0,D} M_B(< r_B) \langle r^2_B \rangle, 
\end{split} 
\end{equation}
\noindent  Note that the integral above is effectively over the baryonic extent, i.e.,  a limited subsystem within the DM halo. The baryonic average square radius above is defined by:
\begin{equation}
\langle r^2_B \rangle \equiv 
{\int_0^{r_B} 4\pi r^4  \rho_B(r) dr \over \int_0^{r_B} 4\pi r^2 \rho_B(r) dr} = 
{\int_0^{r_B} r^2 \rho_B(r) dV \over M_B (< r_B)}. \label{eq:r2b}
\end{equation}
\noindent Inserting the potential energy equations (Eqs. \ref{eq:WB} and \ref{eq:WBDappr}) into Eq. \ref{eq:2VT_v2_WM} leads to the 2VT formulation:
\begin{equation}
\langle v^2 \rangle _{\rm 2VT}
   = \langle v^2_B \rangle + \frac{4 \pi G}{3} \rho_{0,D} \mathcal{R} r_B^2, 
   \label{eq:2VT_v2}
\end{equation}
where the dimensionless parameter $\mathcal{R}$ is defined by:
\begin{equation}
\mathcal{R} \equiv {\langle r^2_B \rangle \over r_B^2}. \label{eq:R_param}
\end{equation}
Note that if the baryons are more concentrated ($\mathcal{R} \ll 1$), the DM correction is weak.
If the baryons are more extended ($\mathcal{R}\gtrsim 1$), the 2VT correction is proportionally stronger.

Now,  Eq. \ref{eq:2VT_v2} can be expressed in terms of accelerations at the gravitational radius of the baryonic component, $r_B$ (c.f. Eq. \ref{eq:grG}). Using Eq. \ref{eq:gr_rhoconst} with $\rho = \rho_{0,D} \approx {\rm const.}$, then, {\it the 2VT in terms of accelerations} is written as:

\begin{equation}
g_{\rm 2VT}(r_B) = g_{\rm vir,B} (r_B) + \mathcal{R}g_D(r_B), \label{eq:2VT_g}
\end{equation}
\noindent where:
\begin{equation}
g_{\rm vir,B}(r_B) = {\langle v^2_B \rangle \over r_B},   \label{eq:gB}
\end{equation}
\begin{equation}
g_D(r_B) = {4 \pi G\over 3}  r_B  \rho_{0,D} ,  \label{eq:gD}
\end{equation}
where $g_D(r_B)$ is an effective DM acceleration at $r_B$ (from the constant-density approximation).
The 2VT relation (Eq. \ref{eq:2VT_g}) should not be interpreted as a local force law, but rather as a global identity evaluated at the baryonic gravitational radius. Therefore, the 2VT defines an effective acceleration scale associated with the dynamical equilibrium of the system, which can be formally interpreted as the acceleration experienced by a test mass at $r_B$. This scale is hence associated with a baryonic system embedded in an extended DM halo by the addition (to the 1VT expression) of a {\it geometrically baryon-weighted contribution of the DM halo}, given by the term $\mathcal{R}g_D$. Accelerations in the 2VT relation are global and set at the baryonic gravitational radius.

\subsection{The Constant-Interaction Approximation Model, CIA} \label{ssec:CIA}

The relation in Eq.~\ref{eq:2VT_g} is strictly defined at the baryonic gravitational radius $r_B$, where the virial theorem applies. As such, it provides a global constraint on the dynamical state of the baryonic component embedded in the DM halo, and defines a characteristic acceleration scale associated with this coupling, given by the term $\mathcal{R} g_D(r_B)$. However, the RAR is a radially resolved relation, involving measurements at different radii within galaxies. The virial theorem, being a global relation, does not by itself determine the radial dependence of the acceleration. In particular, Eq.~\ref{eq:2VT_g} cannot be directly extended to arbitrary radii without additional assumptions.

We emphasize that, throughout this work, $g_B(r)$ denotes the local Newtonian acceleration inferred from the enclosed baryonic mass: $g_B= G M_B(<r_{obs})/r^2_{obs}$, while in the 2VT relation we use $g_{\rm vir, B}(r_B)$ instead (cf. Eqs. \ref{eq:2VT_g} and  \ref{eq:gB}), formally defining a global quantity evaluated at the baryonic gravitational radius. 
In this sense, the CIA does not equate $g_{\rm vir, B}(r_B)$ with  $g_B(r)$, but rather replaces the global virial relation by a local Newtonian description of the baryonic component, while retaining the DM contribution as an effective interaction term.

To establish a connection with the RAR, we note that the 2VT derivation assumes that, within the region occupied by the baryonic component, the DM halo can be approximated by a distribution with nearly constant density, $\rho_{0,D}$, (c.f. Eq. \ref{eq:WBDappr}). Under this condition, the DM contribution to the gravitational acceleration behaves as
\begin{equation}
g_D(r) = \frac{G M_D(<r)}{r^2} \propto r,
\end{equation}
and therefore varies on a radial scale significantly larger than that over which $g_B(r)$ changes.

In contrast, the baryonic acceleration $g_B(r)$ can exhibit a much stronger radial variation, reflecting the typically more concentrated distribution of baryons. As a consequence, over the radial range probed by the RAR within a given system, the DM contribution $g_D(r)$ varies slowly compared to the baryonic acceleration, which typically changes by orders of magnitude across the same radial range. As a result, the DM contribution can be approximated to first order as an effective constant.
This approximation applies at the level of individual systems, where measurements at different radii sample a limited region of the DM halo. At the same time, when considering a population of galaxies, the corresponding effective value may vary from system to system, reflecting differences in their global structural properties. Within this framework, the CIA prediction for the radial acceleration relation at a given radius, $r_{\mathrm{obs}}$, is then given by
\begin{equation}
g_{\rm CIA}(r_{\mathrm{obs}}) \simeq g_B(r_{\mathrm{obs}}) + g_{\rm int},
\end{equation}
where the effective interaction term $g_{\rm int}$ encapsulates the contribution of the DM halo within the baryonic region for a given system.  The 2VT then provides a natural estimate for the effective interaction term, namely
\begin{equation}
g_{\rm int} \equiv \mathcal{R}\, g_D(r_B),
\end{equation}
which is fixed by the global structure of the system through the baryonic mass distribution and the mean DM density within $r_B$. In this sense, the approximation of a radially constant DM contribution is applied locally within each galaxy, while the 2VT fixes the characteristic scale of this contribution across the population.

In this formulation, the 2VT does not predict the detailed radial dependence of the acceleration, but instead sets the scale of an effective contribution arising from the embedding of the baryonic component within the DM halo. The observed RAR can then be interpreted as the result of combining a local baryonic acceleration, governed by Newtonian gravity, with a slowly varying DM contribution whose characteristic amplitude is fixed by the 2VT.

This interpretation makes explicit the distinction between global dynamical constraints and local gravitational relations. The virial theorem (in both its one- and two-component forms) determines global properties of the system, while the radial dependence of the acceleration follows from the underlying mass distribution. The success of the above approximation in reproducing the RAR indicates that, within the baryonic regions of galaxies, the DM contribution can be effectively represented by a single acceleration scale, closely related to that defined by the 2VT.


\subsection{Logarithmic representation of the CIA and its limiting behavior} \label{ssec:log_CIA}

Throughout this work, $\log$ denotes the base-10 logarithm, unless explicitly indicated otherwise (e.g., in figure labels where $\log_{10}$ is retained for clarity). Starting from the CIA relation written in terms of accelerations (Eqs. \ref{eq:CIA} and \ref{eq:CIA_int}), one sees that the DM contribution enters as a simple additive term in linear space.
However, the RAR is usually presented in logarithmic variables,
$(\log g_B, \log g_{\rm CIA})$. Taking the logarithm of the CIA relation yields
\begin{equation}
\log g_{\rm CIA} = \log\left(g_B + g_{\rm int} \right).
\end{equation}
\noindent In this form, the second term cannot be isolated as a separate additive term ($\log(a+b) \neq \log a + \log b$). A more informative expression is obtained by factoring out $g_B$,
\begin{equation}
\log g_{\rm CIA}
=
\log g_B
+
\log\left(1+\frac{g_{\rm int}}{g_B}\right).
\end{equation}
\noindent This shows that in logarithmic space the effective correction term is
\begin{equation}
\Delta_{\log} =
\log\left(1+\frac{g_{\rm int}}{g_B}\right),
\end{equation}
which is a dimensionless correction term governing the deviation from the Newtonian relation in logarithmic space. The constant interaction term therefore does not appear as a constant vertical shift in the $(\log g_B,\log g_{\rm CIA})$ plane, but rather as a nonlinear term that modifies the curvature of the relation. Two limiting regimes clarify this behavior. When the baryonic acceleration dominates ($g_B \gg g_{\rm int}$), one has
\begin{equation}
\log g_{\rm CIA}
\approx
\log g_B ,
\end{equation}
so that the relation approaches the Newtonian relation, $g_{\rm CIA} \simeq g_B$. Conversely, when the dark component dominates, ($g_B \ll g_{\rm int}$), the observed acceleration becomes approximately constant,
$g_{\rm CIA} \approx g_{\rm int}$, and therefore
\begin{equation}
\log g_{\rm CIA} \approx \log g_{\rm int},
\end{equation}
which corresponds to a horizontal asymptote in the $(\log g_B,\log g_{\rm CIA})$ plane. Thus, while the CIA corresponds to a simple linear sum of accelerations in physical space, in the logarithmic representation used for the RAR the dark component manifests itself as a curvature controlled by the dimensionless ratio $g_{\rm int}/ g_B$ rather than as a simple additive offset.

\subsection{The Virial-Motivated Interaction Model, VIM} \label{APP_VIM}

Starting from the definition of the baryon–DM interaction term in the 2VT (Eq. \ref{eq:WBD}),
\begin{equation}
W_{BD}
= - G \int_0^\infty \frac{\rho_B(r) M_D(<r)}{r} dV ,
\label{eq:WBD_copy}
\end{equation}
\noindent where the enclosed dark-matter mass is given by
\begin{equation}
M_D(<r) = 4\pi \int_0^r \rho_D(r') r'^2  dr' ,
\label{eq:DM_Mass}
\end{equation}
one may re-express the interaction energy as a nested radial integral.

Assuming spherical symmetry, we define: 
\begin{equation}
\alpha(r) \equiv \frac{\rho_B(r)}{\rho_D(r)}.
\label{eq:alphaR}
\end{equation}
\noindent Then, Eq. \eqref{eq:WBD_copy} becomes
\begin{equation}
W_{BD}
= - 16\pi^2  G
\int_0^\infty \alpha(r)  r  \rho_D(r) dr
\int_0^r    \rho_D(r') r'^2  dr'.
\label{eq:WBD_nested}
\end{equation}

\noindent In this formulation, the outer integral corresponds to the contribution of baryonic mass elements located at radius $r$, while the inner integral represents the dark-matter mass enclosed within that radius. By virtue of spherical symmetry and Newton's shell theorem, only the dark-matter mass interior to a given baryonic shell contributes to the gravitational interaction, which naturally leads to the nested structure of Eq. \eqref{eq:WBD_nested}.

\noindent Now, adopting a proportionality hypothesis in Eq. \ref{eq:alphaR}, we write:
\begin{equation}
\alpha(r) \equiv \alpha \doteq ~{\rm const.},
\label{eq:alpha}
\end{equation}
\noindent where the symbol $\doteq$ does not indicate a literal proportionality of the underlying density profiles but rather a placeholder for an effective, scale-dependent parametrization of the baryon–DM coupling.

It is then meaningful to introduce a {\it cumulative interaction energy}, defined by truncating the outer integral at a finite observational (deprojected) baryonic radius, $r_{\mathrm{obs}}$,
\begin{equation}
W_{BD}(<r_{\mathrm{obs}})
\equiv
- 16\pi^2 \alpha G
  \int_0^{r_{\mathrm{obs}}}  r  \rho_D(r) dr 
  \int_0^r   \rho_D(r') r'^2  dr'  .
  \label{eq:WBD_cumulative}
  \end{equation}
\noindent 
This quantity represents the cumulative gravitational interaction energy between the baryonic component enclosed within radius $r_{\mathrm{obs}}$ and the DM mass interior to each baryonic shell. The truncation at $r_{\mathrm{obs}}$ reflects the finite spatial extent of the observed baryonic system and explicitly incorporates the radial structure of the DM halo. As such, $W_{BD}(<r_{\mathrm{obs}})$ provides a scale–dependent measure of the baryon–DM gravitational coupling within finite apertures.

In contrast with the CIA, where the baryon--DM coupling is represented by an effective constant interaction term $g_{\rm int}$, the cumulative quantity $W_{BD}(<r_{\mathrm{obs}})$ retains explicit information about the radial structure of the DM halo. It therefore provides the natural starting point for a fully local description of the interaction.

Although $W_{BD}(<r_{\mathrm{obs}})$ has dimensions of energy, it cannot be directly converted to a local acceleration using Eqs. \ref{eq:grG} and \ref{eq:2VT_v2_WM}. To obtain a physically meaningful local acceleration associated with the baryon–dark–matter interaction, the cumulative energy must first be localized. This is achieved by considering the radial derivative of the cumulative interaction energy, which isolates the contribution of the shell at radius $r$. We therefore define the {\it local interaction acceleration} as
\begin{equation}
g_{BD}(r)
\equiv
\frac{1}{M_B(<r)}
\frac{1}{4\pi r^2}
\left|
\frac{d W_{BD}(<r)}{dr}
\right| .
\label{eq:gBD_definition}
\end{equation}
The normalization by $M_B(<r)$ ensures that the resulting quantity represents an acceleration per unit baryonic mass enclosed within the radius, consistent with the global normalization adopted in the virial formulation. The derivative converts the cumulative quantity into a shell–level response, while the factor $1/(4\pi r^2)$ accounts for the geometric dilution of this response over a spherical surface. This step is purely geometrical and does not rely on identifying $r$ with a gravitational radius. From Eqs. \ref{eq:DM_Mass} and \ref{eq:WBD_cumulative} we find that the derivative can be evaluated analytically, yielding
\begin{equation}
g_{BD}(r)
=
\frac{4\pi \alpha G}{M_B(<r)}
\frac{\rho_D(r)\, M_D(<r)}{r}.
\label{eq:gBD_local}
\end{equation}

Equation~\eqref{eq:gBD_local} provides a local, radius–dependent expression for the interaction contribution to the acceleration. In the limit where $g_{BD}(r)$ varies slowly over the baryonic region, this expression reduces effectively to a constant interaction term, thereby recovering the CIA description. In general, however, $g_{BD}(r)$ retains its explicit radial dependence.

The VIM prediction for the radial acceleration relation at a given radius, $r_{\mathrm{obs}}$, is then given by
\begin{equation}
g_{\rm VIM} (r_{\mathrm{obs}})  = g_B (r_{\mathrm{obs}}) + g_{BD}(r_{\mathrm{obs}}),\label{eq:g_ext}
\end{equation}
where $g_{BD}(r_{\mathrm{obs}})$ refers to the numerically computed interaction contribution (Eq. \ref{eq:gBD_local}).

\subsection{Dark matter density profiles for the VIM} \label{APP_profiles}

Here we provide a brief compendium of formulae for the four DM profiles used in the present work. These profiles span cuspy (NFW, Moore) and cored (Burkert, cored isothermal) behaviors, allowing us to probe the sensitivity of the VIM predictions to the inner structure of the DM halo.

{\sl (1) Navarro–Frenk–White, NFW } \citep{Nav96,Nav97}. \\

\noindent - 3D mass density profile:
$$
\rho_{\mathrm{NFW}}(r) = \frac{\rho_0}{\tfrac{r}{r_s}\,(1 + r/r_s)^2}.
$$

{\sl (2) Burkert} \citep{Bur95}. \\

\noindent  - 3D mass density profile:
$$
\rho_{\mathrm{Bur}}(r) = \frac{\rho_0}{(1 + r/r_s)(1+(r/r_s)^2)}.
$$

{\sl (3) Cored Isothermal} \citep{Bin87}. \\

\noindent - 3D mass density profile:
$$
\rho(r)_{\rm coreISO} = \frac{\rho_0}{1+(r/r_s)^2}.
$$

{\sl (4) Moore et al.} \citep{Moo99}.

\noindent - 3D mass density profile:
$$
\rho_{\mathrm{Moore}}(r) = \frac{\rho_0}{(r/r_s)^{1.5}\,(1+r/r_s)^{1.5}}.
$$

\bibliographystyle{aasjournal}
\bibliography{References}

\end{document}